\documentclass[11pt]{article}
\usepackage{axodraw}
\usepackage{epsfig}
\usepackage{amsfonts}
\usepackage{amsmath}
\usepackage{bbm}
\input pix.sty

\newcommand{\ir}{\rmii{UV}} 
\renewcommand{\eq}{Eq.~}
\renewcommand{\eqs}{Eqs.~}
\renewcommand{\se}{Sec.~}

\renewcommand{\fig}{Fig.~}

\newcommand{\mD}{m_\rmi{D}}

\newcommand{\alphas}{\alpha_{\rm s}}
\newcommand{\Nf}{N_{\rm f}}
\newcommand{\Nc}{N_{\rm c}}

\newcommand{\rmO}{{\mathcal{O}}}
\newcommand{\bmu}{\bar\mu}

\def\lsi{\raise0.3ex\hbox{$<$\kern-0.75em\raise-1.1ex\hbox{$\sim$}}}
\def\gsi{\raise0.3ex\hbox{$>$\kern-0.75em\raise-1.1ex\hbox{$\sim$}}}
\newcommand{\lsim}{\;\lsi\;} 
\newcommand{\gsim}{\;\gsi\;} 

\newcommand{\nB}{n_\rmii{B}}

 \renewcommand{\nB}[1]{n_\rmii{B{#1}}}
\newcommand{\rmii}[1]{{\mbox{\tiny\rm{#1}}}}
\newcommand{\re}{\mathop{\mbox{Re}}}
\newcommand{\im}{\mathop{\mbox{Im}}}

\newcommand{\Tint}[1]{{\hbox{$\sum$}\!\!\!\!\!\!\!\int\,}_{\!\!\!\!\raise-0.9ex\hbox{$\scriptstyle{#1}$}}}
\newcommand{\Tinti}[1]{{{\Sigma}\!\!\!\!\raise0.3ex\hbox{$\int$}_\rmii{${#1}$}}}

\newcommand{\bi}{\begin{itemize}}
\newcommand{\ei}{\end{itemize}}

\newcommand{\hide}[1]{ }

\def\TAsc(#1,#2)(#3,#4,#5)%
{\SetWidth{2.0}\CArc(#1,#2)(#3,#4,#5)\SetWidth{1.0}}
\def\Lwidth{3}

\def\TAgl(#1,#2)(#3,#4,#5){\SetWidth{2.0}\PhotonArc(#1,#2)(#3,#4,#5){\Lwidth}%
{6.283 #3 mul 360 div #4 #5 sub #4 #5 sub mul sqrt mul Tdensity mul}%
\SetWidth{1.0}}
\def\TLgl(#1,#2)(#3,#4){\SetWidth{2.0}\Photon(#1,#2)(#3,#4){\Lwidth}
{#1 #3 sub #1 #3 sub mul #2 #4 sub #2 #4 sub mul add sqrt Tdensity mul}%
\SetWidth{1.0}}
\newcommand{\piC}[1]{\;\parbox[c]{40pt}{\begin{picture}(120,60)(0,-20)
\SetWidth{1.0}\SetScale{0.35} #1 \end{picture}}\;}
\def\ConnectedA(#1,#2,#3){\piC{#1(60,-15)(75,34,146) #2(60,75)(75,214,326)%
 #3(60,60)(20,190,350)%
 \GBoxc(0,30)(10,10){1} \GBoxc(120,30)(10,10){1}%
  }}
\def\ConnectedB(#1,#2,#3){\piC{#1(60,-15)(75,34,146) #2(60,75)(75,214,326)%
 #3(60,60)(60,0)%
 \GBoxc(0,30)(10,10){1} \GBoxc(120,30)(10,10){1}%
  }}
\def\ConnectedC(#1,#2){\piC{#1(60,-15)(75,34,146) #2(60,75)(75,214,326)%
 \GBoxc(0,30)(10,10){1} \GBoxc(120,30)(10,10){1}%
  }}
\def\ConnectedD(#1,#2){\piC{#1(60,-15)(75,34,146) #2(60,75)(75,214,326)%
 \GBoxc(0,30)(10,10){1} \GBoxc(120,30)(10,10){1}%
 \SetWidth{2.0} 
 \Line(55,55)(65,65)%
 \Line(55,65)(65,55)
  }}
\newcommand{\piG}[1]{\;\parbox[c]{300pt}{\begin{picture}(150,40)(-175,-15)
\SetWidth{1.0}\SetScale{1.0} #1 \end{picture}}\;}
\def\LattGE{\piG{%
 \SetWidth{1.0} 
 \Line(-130,10)(-110,10)%
 \Line(-100,10)(-100,20)\Line(-100,20)(-90,20)%
 \Line(-85,10)(-75,10)\Line(-75,10)(-75,20)%
 \Line(-65,20)(-45,20)%
 \Line(-35,20)(-35,10)\Line(-35,10)(-25,10)%
 \Line(-20,20)(-10,20)\Line(-10,20)(-10,10)%
 \Vertex(-130,10){1}
 \Vertex(-110,10){1}
 \Vertex(-100,10){1}
 \Vertex(-90,20){1}
 \Vertex(-85,10){1}
 \Vertex(-75,20){1}
 \Vertex(-65,20){1}
 \Vertex(-45,20){1}
 \Vertex(-35,20){1}
 \Vertex(-25,10){1}
 \Vertex(-20,20){1}
 \Vertex(-10,10){1}
 \Text(-87.5,15)[c]{$-$}
 \Text(-22.5,15)[c]{$-$}
 \Text(-135,15)[c]{$\bigl\langle\,$}
 \Text(-105,15)[c]{$\bigl($}
 \Text(-70,15)[c]{$\bigr)$}
 \Text(-40,15)[c]{$\bigl($}
 \Text(-5,15)[c]{$\bigr)$}
 \Text(2,15)[c]{\small $+$}
 \Line(5,20)(25,20)%
 \Line(35,20)(35,10)\Line(35,10)(45,10)%
 \Line(50,20)(60,20)\Line(60,20)(60,10)%
 \Line(70,10)(90,10)%
 \Line(100,10)(100,20)\Line(100,20)(110,20)%
 \Line(115,10)(125,10)\Line(125,10)(125,20)%
 \Vertex(5,20){1}
 \Vertex(25,20){1}
 \Vertex(35,20){1}
 \Vertex(45,10){1}
 \Vertex(50,20){1}
 \Vertex(60,10){1}
 \Vertex(70,10){1}
 \Vertex(90,10){1}
 \Vertex(100,10){1}
 \Vertex(110,20){1}
 \Vertex(115,10){1}
 \Vertex(125,20){1}
 \Text(47.5,15)[c]{$-$}
 \Text(112.5,15)[c]{$-$}
 \Text(135,15)[c]{$\bigr\rangle$}
 \Text(30,15)[c]{$\bigl($}
 \Text(65,15)[c]{$\bigr)$}
 \Text(95,15)[c]{$\bigl($}
 \Text(130,15)[c]{$\bigr)$}
 \Text(-138,15)[r]{$\sum_{i=1}^{3}\re\tr$}
 \Line(-190,3)(140,3)
 \Text(-50,-10)[r]{$- 6 a^4 \re\tr\bigl\langle$}
 \Line(-48,-10)(15,-10)
 \Vertex(-48,-10){1}
 \Vertex(15,-10){1}
 \Text(20,-10)[c]{$\bigr\rangle$}
 \Text(-195,2)[r]{$G_E(\tau) = $}
 \LongArrow(175,-5)(175,15)%
 \LongArrow(175,-5)(155,-5)%
 \Text(157,1)[c]{\small $x_0$}
 \Text(167,13)[c]{\small $x_i$}
  }}

\makeatletter \@addtoreset{equation}{section} \makeatother
\renewcommand{\theequation}{\arabic{section}.\arabic{equation}}
\makeatletter
\renewcommand\section{\@startsection {section}{1}{\z@}%
                                   {-5.5ex \@plus -1ex \@minus -.2ex}
                                   {2.3ex \@plus.2ex}%
                                   {\normalfont\large\bfseries}}
\renewcommand\subsection{\@startsection{subsection}{2}{\z@}%
                                     {-3.25ex\@plus -1ex \@minus -.2ex}%
                                     {1.5ex \@plus .2ex}%
                                     {\normalfont\normalsize\bfseries}}
\renewcommand\thesection {\@arabic\c@section}
\renewcommand\thesubsection   {\thesection.\@arabic\c@subsection}
\renewcommand{\@seccntformat}[1]{%
\csname the#1\endcsname.\hspace{1.0em}}
\makeatother

\begin{document}

\begin{titlepage}
\begin{flushright}
BI-TP 2009/01\\
\vspace*{1cm}
\end{flushright}
\begin{centering}
\vfill

{\Large{\bf
A way to estimate the heavy quark thermalization rate 
\\[2mm]
from the lattice
}} 

\vspace{0.8cm}

Simon~Caron-Huot$^\rmi{a}$, 
Mikko~Laine$^\rmi{b}$, 
Guy D.~Moore$^\rmi{a,b}$ 

\vspace{0.8cm}

$^\rmi{a}$%
{\em
Department of Physics, McGill University, 
Montr\'eal, QC H3A 2T8, Canada\\ }

\vspace*{0.3cm}

$^\rmi{b}$%
{\em
Faculty of Physics, University of Bielefeld, 
D-33501 Bielefeld, Germany\\ }

\vspace*{0.8cm}

\mbox{\bf Abstract}
 
\end{centering}

\vspace*{0.3cm}
 
\noindent
The thermalization rate of a heavy quark is related to
its momentum diffusion coefficient. Starting from a Kubo relation
and using the framework of the heavy quark effective theory, we argue that
in the large-mass limit the momentum diffusion coefficient can be defined
through a certain Euclidean correlation function, involving color-electric 
fields along a Polyakov loop. Furthermore, carrying out a perturbative 
computation, we demonstrate that the spectral function corresponding to 
this correlator is relatively flat at small frequencies. Therefore, 
unlike in the case of several other transport coefficients, for which 
the narrowness of the transport peak makes analytic continuation from 
Euclidean lattice data susceptible to severe systematic uncertainties, 
it appears that the determination of the heavy quark thermalization rate 
could be relatively well under control.

\vfill


\vspace*{1cm}
  
\noindent
February 2009

\vfill

\end{titlepage}

%
\section{Introduction}

One of the very interesting discoveries of the RHIC program at Brookhaven
National Laboratory 
has been that heavy quarks (particularly the charm quarks) appear to 
thermalize just about as effectively as the light quarks. This has been 
inferred from measuring the electron $p_T$-spectra produced by the decays
of the heavy quarks, showing indications of the same type of hydrodynamic
flow as experienced by the lighter quarks~\cite{exp}. 

{}On the theoretical side, the historical starting point for a QCD-based
understanding of the behavior of heavy quarks in a thermal environment
was the determination of their energy loss rate to the leading order
in the QCD coupling constant, $\alphas$~\cite{bt}. The energy loss
is directly related to a number of other concepts, such as the diffusion
and the thermalization rates of the heavy quarks~\cite{bs,mt}. In particular, 
assuming that the effective value of $\alphas$ is relatively small
leads to the thermalization rate 
$
 \Gamma \sim \alphas^2 T^2  / M
$, 
where $T$ is the temperature and $M$ is the heavy quark mass~\cite{mt}.
The comparable thermalization rate for a light quark or gluon
is $\Gamma \sim \alphas^2 T$, or at very high energies
$\Gamma \sim \alphas^2 T \sqrt{T/E}$~\cite{bdps}.  
Hence heavy quarks with $M \gg T$ are
expected to thermalize slowly, particularly at weak coupling.

As already mentioned, the empirical facts appear however to be in 
conflict with a slow thermalization rate~\cite{exp}. This has lead
to a lot of new theoretical ideas, with the hope of bringing the 
theoretical determination of $\Gamma$ beyond the leading order 
in $\alphas$. In particular, possibilities for a lattice determination
were explored~\cite{pt}; computations in a strongly coupled theory with 
some similarities with QCD were carried out~\cite{sea,ct}; phenomenological
model treatments of bound states were considered~\cite{rapp}; and the 
first weak-coupling corrections to the leading order result were
determined~\cite{chm}. The studies~\cite{sea}--\cite{chm} 
showed that there could 
indeed be substantial corrections (of a positive sign)
to the leading order result; at the same
time, the study~\cite{pt} showed that a direct lattice determination of the 
heavy quark related observables would be extremely hard, because the 
physics resides in a ``transport peak'' of a certain spectral function, 
of width $\Delta\omega \sim \Gamma \sim \alphas^2 T^2 / M \ll T$, 
which regime is practically 
impossible to explore with Euclidean techniques.

The purpose of this paper is to reconsider the prospects for
a lattice determination, making use
of heavy quark effective theory~\cite{hqet}
in order to systematically
consider the behavior of the heavy quarks in the limit $M \gg T$.
Essentially, this allows us to restrict the attention to the 
``numerator'' of the thermalization rate, $\sim \alphas^2 T^2$, 
which remains finite in the heavy quark limit. In fact, our
main goal will be to derive an observable measurable on the lattice
which has its structure at ``large'' frequencies, 
$\omega \sim \alphas^\rmii{1/2} T \gg \Gamma$, 
and can be addressed much more easily 
than $\Gamma$ itself. 

We note that, in many respects, our study parallels that in ref.~\cite{ct}. 
The main differences are that we use the imaginary-time formalism rather 
than the real-time one, in order to make contact 
with the Euclidean spacetime accessible to lattice techniques; and that 
we try to keep explicit track of $\rmO(\alphas)$ quantum corrections
and renormalization issues.

The plan of this paper is the following. 
In \se\ref{se:red} we derive, by going through several intermediate steps, 
the observable alluded to above, consisting of 
color-electric fields along a Polyakov loop. 
In \se\ref{se:pert} we analyze the corresponding spectral 
function perturbatively, demonstrating a relatively flat 
behavior at small $\omega \lsim \alphas^\rmii{1/2} T$.
In \se\ref{se:latt} we suggest a lattice discretization for 
the object derived in \se\ref{se:red}, while
\se\ref{se:concl} offers some conclusions and an outlook. 

%
\section{Reduction of the current-current correlator}
\la{se:red}

In order to proceed with our derivation, we focus on one of the heavy
quarks of physical QCD; either the charm or the bottom quark. 
We assume for the moment the use of dimensional regularization 
in order to regulate the theory
(though we do not indicate this explicitly). Then there is only one 
large scale
in the system, namely the (renormalized) heavy quark mass, $M$, and 
the task is to account for its effects analytically in the 
basic observable to be defined presently (\eq\nr{rhoV}). 
In \se\ref{se:latt} we return to the complications 
emerging in lattice regularization.

%
\subsection{Definitions}

The diffusive motion of heavy quarks within a thermalized medium
can be characterized by four different quantities, all of which 
are related to each other (at least in the weak-coupling limit). 
We start by defining the ``diffusion
coefficient'', $D$, proceeding then to the ``relaxation rate'' 
or ``drag coefficient'', $\eta_D$,
and the ``momentum diffusion coefficient'', $\kappa$.
The fourth quantity, the energy loss ${\rm d} E / {\rm d}x$, 
is historically the first one addressed within QCD~\cite{bt}; 
yet it is not obvious how it could be related to the others
on the non-perturbative level, 
so we omit it from the discussion below.   

Among the three quantities, the one that can most directly be
defined within quantum field theory is the diffusion coefficient $D$.
Consider the spectral function related to the current-current correlator, 
\be
 \rho^{\mu\nu}_V (\omega) 
 \equiv 
 \int_{-\infty}^\infty 
 \!\! {\rm d}t \,   e^{i \omega t}
 \!
 \int \! {\rm d}^{3} 
 \vec{x}\,
 \left\langle
  \fr12 {[ 
  \hat {\cal{J}}^\mu(t,\vec{x}), 
  \hat {\cal{J}}^\nu(0,\vec{0})
   ]}
 \right\rangle
 \;, \la{rhoV} 
\ee
where 
$
 \hat {\mathcal{J}}^\mu
 \equiv
 \hat{\bar\psi}\, \gamma^\mu\, \hat\psi  
$; 
$\hat\psi$ is the heavy quark field operator in the Heisenberg picture; 
$
 \langle \ldots \rangle \equiv {\mathcal{Z}^{-1}} 
 \tr[ (...) e^{-\beta\hat H}]
$
is the thermal expectation value; and 
$
 \beta \equiv {1}/{T} 
$
is the inverse temperature. Diffusive motion leads to a pole 
in the spectral function at $\omega = - i D \vec{k}^2$, where $\vec{k}$
is the momentum (already set to zero in \eq\nr{rhoV}). Solving for
the pole position and making use of various symmetries leads 
to the Kubo relation (see, e.g., chapter 6 of ref.~\cite{kg})
\be
 D = \fr1{3 \chi^{00}}\lim_{\omega\to 0} 
  \sum_{i=1}^3 \frac{\rho^{ii}_V(\omega)}{\omega}
 \;. \la{D_def}
\ee
Here $\chi^{00}$ corresponds to a ``susceptibility'' related 
to the conserved charge 
$\int\! {\rm d}^3\vec{x} \, \hat {\cal J}^0(t,\vec{x})$, 
\be
 \chi^{00} \equiv
 \beta \int \! {\rm d}^{3} 
 \vec{x}\,
 \left\langle
  \hat {\cal{J}}^0(t,\vec{x}) \,  
  \hat {\cal{J}}^0(t,\vec{0})
 \right\rangle
 \;. \la{susc_def}
\ee
For a dilute system of heavy quarks, $T\chi^{00}$ defines
their ``number density''\footnote{%
 In the non-relativistic limit and at zero chemical potential, 
 $ 
    T \chi^{00} = 
    4\Nc ( {MT} / {2\pi} )^{3/2} \exp({-\beta M})
 $; 
 however, our basic arguments hold also at a non-zero 
 chemical potential for the heavy quarks,
 whereby the exponential suppression could be 
 removed from $T \chi^{00}$.}.
Note that the conserved vector current $\hat {\cal J}^\mu$
does not require renormalization, so that
the definitions in \eqs\nr{D_def} and \nr{susc_def} are guaranteed 
to be ultraviolet finite at any order.

To define the other quantities, we need to assume
that the spectral function around zero frequency possesses
a narrow transport peak.
Due to a heavy quark's large inertia, this
is certainly true for $M$ sufficiently large,
which we assume to be the case from now on.
In this limit, 
the spectral function will 
on general grounds take the form of a Lorentzian\footnote{%
 Two concrete examples for how such a dependence 
 on $\omega$ can arise are reviewed in appendix A.
 },
\be
 \sum_i 
 \frac{\rho^{ii}_V (\omega)}{\omega}
 \; \stackrel{\;\;\;\omega \lsim \omega_\ir }{=} \; 
 3\chi^{00} D \frac{\eta_D^2}{\eta_D^2+\omega^2}
 \;, \la{lorentz}
\ee
where $\omega_\ir$ is a frequency scale at which
the Lorentzian is overtaken by other types of physical processes.

The other two transport coefficients are then defined
from the properties of the transport peak.
We define the ``drag coefficient'' $\eta_D$ to be the width
of the Lorentzian, and the (a priori mass-dependent)
``momentum diffusion coefficient''
$\kappa^{(M)}$ to be $M_\rmi{kin}^2$ times
the coefficient of the power-law falloff of its tails,
\be
 \kappa^{(M)} \equiv \frac{M_\rmi{kin}^2\omega^2 }{3T\chi^{00}}
              \left.\sum_i \frac{2 T \rho^{ii}_V(\omega)}{\omega}
              \right|^{\;\eta_D\,\ll \;|\omega|\, \lsim \omega_\ir}\;.
   \la{kappa0}
\ee
Here $M_\rmi{kin}$ refers to the heavy quark's kinetic mass,
to be defined presently (cf.\ \eq\nr{defM}).  Later on we will
define a transport coefficient $\kappa$ from the $M\to\infty$
limit of $\kappa^{(M)}$.

The physical motivation for the definition
in \eq\nr{kappa0} is as follows.
In the dilute limit the current $\hat{\cal{J}}^\mu$ couples
individually to the heavy quarks; the spectral function
$\rho^{\mu\nu}_V(\omega)$ is thus a product of their number 
density $T\chi^{00}$ times a contribution from one heavy quark.
For a single quark, $\int {\rm d}^3\vec{x}\, \hat{\cal{J}}^i \equiv \hat{v}^i$
represents a non-perturbative measurement of its velocity.
Recalling Newton's law, $M_\rmi{kin}{\rm d}\hat{\cal{J}}^i /{\rm d}t$
is the force acting on the heavy quark; thus $\kappa^{(M)}$ 
is a correlator of that force with itself at different times, 
transformed into frequency space. The factor $2 T/\omega$ relates
the spectral function to a time-symmetric correlator, for which 
this classically motivated argumentation applies.
Thus \eq\nr{kappa0} generalizes, in a non-perturbative way, the
force-force correlator definition of $\kappa$ given in ref.~\cite{ct}.
The condition on $\omega$ instructs us to integrate this
force over a time scale, 
long compared with the medium's correlation time
(set by $t \sim \omega_\ir^{-1}$), but
short compared with 
the dynamics of the heavy quark
(set by $t\sim \eta_D^{-1}$).



The coefficients $D$, $\eta_D$ and $\kappa^{(M)}$ thus defined
are related by fluctuation-dissipation relations, which follow
from the fact that the area under the transport peak
defines the (coarse-grained)
equal-time mean-squared velocity of a heavy quark,
\be
  \langle {\bf v}^2 \rangle \equiv
  \frac{1}{T \chi^{00}}\int_{-\infty}^{\infty}
  \frac{{\rm d}\omega}{2\pi}
  \sum_i \frac{2 T \rho^{ii}_V(\omega)}{\omega} G_\ir(\omega) 
  \;.
  \la{area}
\ee
Here we have introduced a cutoff function
$G_\ir(\omega)$ designed to isolate 
the transport peak from other types 
of physics, for instance 
$G_\ir(\omega)=\theta(\omega_\ir-|\omega|)$.
In the time domain we are thus
averaging 
over a time scale $t_\ir \gsim \omega_\ir^{-1}$.
Such a time averaging is mandatory to
make $\langle {\bf v}^2\rangle$ finite and
well-defined, since an
instantaneous measurement of the heavy quark's velocity would
induce it to radiate (or absorb) large amounts of energy,
thereby changing its state.
We emphasize that, were there
no sharp zero-frequency peak in the spectral
function, there would be no unambiguous notion of a heavy quark's
(coarse-grained) mean squared velocity.

Motivated by the standard non-relativistic 
thermodynamic result,
we can now define a kinetic mass via
\be
  \langle {\bf v}^2\rangle\equiv 3\frac{T}{M_\rmi{kin}}\,.
  \la{defM}
\ee
Eqs.~\nr{lorentz}--\nr{area} then yield
the fluctuation-dissipation, or Einstein, relations:
\be
 D=\frac{2T^2}{\kappa^{(M)}}
   \;,\qquad
 \eta_D=\frac{\kappa^{(M)}}{2M_\rmi{kin}T} \;,
 \la{einst}
\ee
both of which involve 
$\rmO({\eta_D}/{\omega_\ir})$
relative uncertainties.
Note that $\eta_D\sim 1/M_\rmi{kin}$ in the large mass limit, assuming 
that $\kappa^{(M)}$ contains no (power-like) dependence on
$M_\rmi{kin}$,
justifying the narrow peak assumption.

Thermodynamic considerations relate
the kinetic mass defined in \eq\nr{defM} to the
standard notion.
Namely, thanks to the slow dynamics
of a heavy quark, one can (approximately)
define a free energy $F(\vec{v})$ as a function of its
velocity (time-averaged over a period $\sim t_\ir$);
expanding it as
$F(\vec{v})=M_\rmi{rest}+M_\rmi{kin}{\vec{v}^2}/{2}+\rmO(\vec{v}^4)$
at small $\vec{v}$ and taking a thermodynamic
average should reproduce \eq\nr{defM}.


The only approximations we have made so far concern
the assumption of a narrow transport peak.
Parametrically, in weakly coupled QCD \cite{mt},
$\rho^{ii}_V(\omega)/(\chi^{00}\omega)$ has a peak value
$D\sim 1/g^4T$, where $g^2 \equiv 4 \pi \alphas$; 
a width $\eta_D\sim g^4T^2/M$; 
and a perturbative ultraviolet contribution 
which will start to depart from the $1/\omega^2$ Lorentzian 
tail at the scale $\omega_\ir \sim gT$ 
(see \se\ref{se:pert}).
Thus errors are of order
$\eta_D/gT\sim g^3T/M$.
In strongly coupled multicolor ($\Nc\to\infty$)
$\mathcal{N}=4$ Super-Yang-Mills theory \cite{sea},
with a 't Hooft coupling $\lambda=g^2\Nc$,
the width of the transport peak
is $\eta_D\sim \sqrt{\lambda}T^2/M$, and the continuum
takes over at $\omega_\ir\sim T$;
thus ambiguities are suppressed by $\sqrt{\lambda}T/M$.

Expecting the force-force correlator
$\kappa^{(M)}$ 
to actually be mass-independent at large $M_\rmi{kin}$,
as will be verified {\it a posteriori}, 
we are finally
led to take the $M_\rmi{kin}\to\infty$ limit of \eq\nr{kappa0},
inside of which it is essential to retain $\omega$ 
small but non-zero:
\be
 \kappa \;\equiv\; 
 \frac{\beta}{3} \sum_{i=1}^{3}  
 \lim_{\omega\to 0}
  \omega^2 \biggl[
  \lim_{M\to\infty} \frac{M_\rmi{kin}^2}{\chi^{00}} 
  \int_{-\infty}^\infty 
 \!\! {\rm d}t \,  e^{i \omega t}
 \! 
 \int \! {\rm d}^{3} 
 \vec{x}\,
 \left\langle
  \fr12 {\Bigl\{ 
  \hat {\cal{J}}^i(t,\vec{x}), 
  \hat {\cal{J}}^i(0,\vec{0})
   \Bigr\} }
 \right\rangle
 \biggr]
 \;. \la{kappa_def}
\ee
The factor $2T/\omega$ has been accounted for by replacing
the spectral function by a time-symmetric correlator. 
Eq.~\nr{kappa_def}
will be the starting point for the further steps to be taken. 


%
%

%
\subsection{Heavy quark limit}

Starting from the definition in \eq\nr{kappa_def}, 
our next goal is to carry out the limit $M\to \infty$. 
As a first step we note that, making use of time translational 
invariance and carrying out partial integrations, 
the definition in \eq\nr{kappa_def} can be rephrased as 
\be
 \kappa =
 \frac{\beta}{3}  \sum_{i=1}^{3} 
 \lim_{\omega\to 0}
  \biggl[
  \lim_{M\to\infty} \frac{M_\rmi{kin}^2}{\chi^{00}}
  \int_{-\infty}^\infty 
 \!\! {\rm d}t \,
  e^{i \omega (t-t')} \!
 \int \! {\rm d}^{3} 
 \vec{x} \,
 \biggl\langle
  \fr12 \biggl\{ 
  \frac{{\rm d} \hat {\cal{J}}^i(t,\vec{x}) }{{\rm d}t}, 
  \frac{{\rm d} \hat {\cal{J}}^i(t',\vec{0})}{{\rm d}t'}
   \biggr\} 
 \biggr\rangle
 \biggr]
 \;. \la{kappa_def_2}
\ee
In order to evaluate the time derivatives here, let us rewrite
the QCD Lagrangian, 
$
 \mathcal{L}_\rmi{\,QCD} = \bar\psi (i \gamma^\mu D_\mu - M) \psi + 
 \mathcal{L}_\rmi{\,light}
$, after a Foldy-Wouthuysen transformation~\cite{kt}:
%
expanding in $1/M$ and dropping total derivatives, this yields 
\ba
 \mathcal{L}_\rmi{\,QCD} & = &   \theta^\dagger\left(
 iD_0 - M + \frac{c_2\, \vec{D}^2 + c_B\, \sigma\cdot g\vec{B} }{2 M} 
 \right)\theta
 + \phi^\dagger\left(
 iD_0 + M - \frac{c_2\, \vec{D}^2 + c_B\, \sigma\cdot g\vec{B}}{2 M}  
 \right)\phi  \nn & + &  \frac{i\, c_E }{2 M} 
 \left( \theta^\dagger \sigma\cdot g\vec{E} \, \phi - 
 \phi^\dagger \sigma\cdot g\vec{E} \, \theta \right) + 
 \rmO\left(\frac{1}{M^2} \right)
 +   \mathcal{L}_\rmi{\,light}
 \;, \la{1oM}
\ea
where $D_i = \partial_i - i g A_i$, 
$g B_i \equiv \fr{i}2 \epsilon_{ijk}[D_j,D_k]$, 
$g E_i \equiv i [D_0,D_i]$,
and $\theta, \phi$ are two-component spinors. 
The mass $M$ is the pole mass%
 \footnote{
   This is true in schemes 
   producing no additive mass renormalization,
   such as dimensional regularization.  There is no
   multiplicative renormalization to $M$ in \eq\nr{1oM} either, 
   because $M$ could be shifted to zero by the field redefinitions
   $\theta\to e^{-iMt}\theta$, $\phi\to e^{iMt}\phi$ 
   and would then remain zero quantum mechanically.
 },
\be
  M = m(\bmu)
 \biggl\{ 1  + \frac{3 g^2 C_F}{(4\pi)^2} 
 \biggl[ \ln\frac{\bmu^2}{m^2(\bmu)} + \fr43 \biggr] 
 + \rmO(g^4)
 \biggr\}
 \;, 
\ee 
where $m(\bmu)$ is the $\msbar$ mass.
In regularization
schemes respecting Lorentz invariance, the coefficient $c_2$ must equal 
unity
(because the combination needed for solving for the pole mass
is $\sim p_0^2 - \vec{p}^2 - M^2$), 
and we assume this to be the case in the following.
The matching
coefficients $c_B, c_E$ equal unity at leading order but have quantum 
corrections; these are not needed in the present study. Note that
the linearly appearing ``rest mass'' $M$ 
is normally shifted away (or rather 
replaced with $0^+$); however, we prefer to keep it explicit for the moment, 
because the shifts needed are non-trivial at a non-zero temperature, 
where the Euclidean time extent is finite.  

Setting $c_2 = 1$ in \eq\nr{1oM}, we can read off the conserved Noether
current and the Hamiltonian in the heavy quark-mass limit: 
\ba
 \hat{\cal J}^0 & = &
 \hat\theta^\dagger\hat\theta
 +
 \hat\phi^\dagger\hat\phi
 \;, \la{cJ00} \\ 
 \hat{\cal J}^j & = &
 \frac{i}{2 M}
 \Bigl[\hat\theta^\dagger (\overleftarrow{\!D}^{\!j} 
 - \overrightarrow{\!D}^{\!j}) \hat\theta - 
 \hat\phi^\dagger (\overleftarrow{\!D}^{\!j} 
 - \overrightarrow{\!D}^{\!j}) \hat\phi \Bigr] +
 \rmO\Bigl(\frac{1}{M^2}\Bigr)
 \;, \la{cJj} \\ 
 \hat H & = & 
 \int \! {\rm d}^3\vec{x} \, 
 \Bigl[
    \hat\theta^\dagger (- gA_0 + M )\hat\theta
  - \hat\phi^\dagger (gA_0 + M )\hat\phi
 \Bigr]+
 \rmO\Bigl(\frac{1}{M}\Bigr)
 \;. 
\ea
Here we treat the fermionic fields as operators but the gauge fields as
c-numbers, anticipating a path integral treatment of the gauge fields.
The time derivatives needed for \eq\nr{kappa_def_2} can subsequently 
be taken according to the canonical equations of motion, 
\be
 \frac{{\rm d}\hat{\cal J}^i}{{\rm d}t}
 = i \bigl[\hat H, \hat{\cal J}^i \bigr] + 
 \frac{\partial\hat{\cal J}^i}{\partial t}
 \;, \la{eom}
\ee
where the partial derivative operates on the background gauge fields. 
The commutator is readily evaluated with the help of equal-time
anticommutators, 
and we also note that since \eq\nr{kappa_def_2} includes a spatial integral
over the currents, partial integrations are allowed. Adding together 
the two parts in \eq\nr{eom} then yields
\be
 \frac{{\rm d}\hat{\cal J}^i}{{\rm d}t}
 = 
 \frac{1}{M} 
 \Bigl\{ 
   \hat\phi^\dagger  g E^i \hat\phi - 
   \hat\theta^\dagger  g E^i \hat\theta 
 \Bigr\} 
 + \rmO\Bigl( \frac{1}{M^2}\Bigr)
 \;.  \la{dJi}
\ee
This can now be inserted into \eq\nr{kappa_def_2}, whereby the explicit
factors of $M$ duly cancel, since $M_{\rm kin}=M$ up to
$\rmO(T/M)$ thermal corrections which vanish in the heavy quark-mass limit:
\be
 \label{eq:kappa_minkowski}
 \kappa = \frac{\beta}{3} \sum_{i=1}^3\lim_{M\rightarrow \infty}
 \frac{1}{\chi^{00}}  \int \! {\rm d}t\, {\rm d}^3 \vec{x}
 \left\langle \fr12 \biggl\{ 
 \left[\hat\phi^\dagger gE^i \hat\phi-
 \hat\theta^\dagger gE^i \hat\theta\right]\!(t,\vec{x})\, ,
 \left[\hat\phi^\dagger gE^i \hat\phi
 -\hat\theta^\dagger gE^i \hat\theta\right]\!(0,\vec{0})
 \biggr\} \right\rangle\,.
\ee
At this point the heavy quarks have become purely static;
the ordering of the limits no longer matters, 
so we have set $\omega\to 0$
inside the Fourier transform. 

Given that our derivation made no use of 
weak-coupling approximations, we believe that \eq\nr{eq:kappa_minkowski}
is free from (even finite) renormalization to all orders in
perturbation theory, in the assumed regularization schemes
with no additive mass renormalization and $c_2$ equal to unity.  
This shows, in particular, that $\kappa$ is $M$-independent.



%
\subsection{Euclidean correlator}

\eq\nr{eq:kappa_minkowski} is a two-point function
of gauge-invariant local operators; it therefore satisfies the standard KMS
conditions which allow us to relate it to a Euclidean correlation function.
In particular, let us define the  Euclidean correlator 
\be
 G_E (\tau) 
 \equiv 
 - \frac{\beta}{3} \sum_{i=1}^{3}
 \lim_{M\to\infty} \frac{1}{\chi^{00}}
 \int\! {\rm d}^3\vec{x}
 \Bigl\langle
  \Bigl[\phi^\dagger gE_i \phi - 
   \theta^\dagger  gE_i \theta\Bigr] (\tau,\vec{x}) \,
  \Bigl[\phi^\dagger gE_i \phi - 
   \theta^\dagger  gE_i \theta\Bigr] (0,\vec{0})
 \Bigr\rangle
 \;. \la{GE}
\ee
Hats have been left out because regular Euclidean path integral techniques
apply for this object, and the minus sign accounts for the fact that  
a Euclidean electric field differs by a factor $i$ from the Minkowskian one. 
The corresponding spectral function can be 
determined by inverting 
(for recent practical recipes see, e.g., 
refs.~\cite{recent}\footnote{%
  The inversion leads to well-known systematic uncertainties, and we 
  have nothing concrete to add on how to treat those. However, as 
  will be demonstrated below, our spectral function is smoother at 
  small $\omega$ than the ones in refs.~\cite{recent}, which should 
  somewhat ameliorate the problems in reconstructing the spectral function.
 }) 
the relation
\be
 G_E(\tau) = 
 \int_0^\infty
 \frac{{\rm d}\omega}{\pi} \rho(\omega)
 \frac{\cosh \left(\frac{\beta}{2} - \tau\right)\omega}
 {\sinh\frac{\beta \omega}{2}} 
 \;, \la{int_rel} 
\ee
or analytically from 
\ba
  \tilde G_E(\omega_n) & \equiv & 
  \int_0^\beta \! {\rm d}\tau \, e^{i \omega_n\tau } G_E(\tau)
  \;, \la{tildeGE} \\ 
  \rho(\omega) & = & \im \tilde G_E(\omega_n \to -i [\omega + i 0^+])
  \;.
 \la{analytic}
\ea
The momentum diffusion coefficient then follows from 
\be
 \kappa = \lim_{\omega\to 0} \frac{2 T}{\omega} \rho(\omega)
 \;. \la{kappa_def_3}
\ee
Note also that by making use of \eq\nr{cJ00}, 
the susceptibility $\chi^{00}$ defined in \eq\nr{susc_def}
can in the Euclidean theory be written as  
\be
 \chi^{00} = \int_0^\beta\! {\rm d}\tau \int\! {\rm d}^3\vec{x}
 \Bigl\langle
  \Bigl[\phi^\dagger \phi + 
   \theta^\dagger \theta\Bigr] (\tau,\vec{x}) \;
  \Bigl[\phi^\dagger  \phi + 
   \theta^\dagger  \theta\Bigr] (0,\vec{0})
 \Bigr\rangle
 \;. \la{susc_def_2}
\ee

In order to work out the contractions 
in \eqs\nr{GE}, \nr{susc_def_2}, we need the heavy quark propagators
within the Euclidean theory
\be
 {\cal L}_E = \theta^\dagger (D_\tau + M )\theta + 
 \phi^\dagger (D_\tau - M) \phi + \rmO\Bigl(\frac{1}{M} \Bigr)
 \;. 
\ee
Making use of the equations of motion satisfied by the propagators, 
together with the proper boundary conditions, 
it can be shown that in the $M\to\infty$ limit
and for $\tau > 0$,  
\ba
 \Bigl\langle \theta_\alpha(\tau,\vec{x}) 
 \, \theta^*_\beta(0,\vec{y})\Bigr\rangle
 & = & 
 \delta^{(3)}(\vec{x-y}) U_{\alpha\beta}(\tau,0)e^{-\tau M}
 \;, \\
 \Bigl\langle \theta_\alpha(0,\vec{x}) 
 \, \theta^*_\beta(\tau,\vec{y})\Bigr\rangle
 & = & 
 - \delta^{(3)}(\vec{x-y}) U_{\alpha\beta}(\beta,\tau)e^{(\tau-\beta) M}
 \;, \\
 \Bigl\langle \phi_\alpha(\tau,\vec{x}) 
 \, \phi^*_\beta(0,\vec{y})\Bigr\rangle
 & = & 
 \delta^{(3)}(\vec{x-y}) U^\dagger_{\alpha\beta}(\beta,\tau)e^{(\tau-\beta) M}
 \;, \\
 \Bigl\langle \phi_\alpha(0,\vec{x}) 
 \, \phi^*_\beta(\tau,\vec{y})\Bigr\rangle
 & = & 
 - \delta^{(3)}(\vec{x-y}) U^\dagger_{\alpha\beta}(\tau,0)e^{-\tau M}
 \;, 
\ea
where $U$ is now a straight fundamental
Wilson line in the Euclidean time direction. 
With these propagators, we obtain 
\ba
 & & \hspace*{-2cm}
 \int\! {\rm d}^3\vec{x}
 \Bigl\langle
  \Bigl[\phi^\dagger gE_i \phi - 
   \theta^\dagger  gE_i \theta\Bigr] (\tau,\vec{x}) \;
  \Bigl[\phi^\dagger gE_i \phi - 
   \theta^\dagger  gE_i \theta\Bigr] (0,\vec{0})
 \Bigr\rangle
 \nn 
 & = & 
 4 \delta^{(3)}(\vec{0}) e^{-\beta M} 
 \Bigl\langle
   \re\tr [U(\beta,\tau)\, gE_i(\tau,\vec{0})
   \, U(\tau,0)\, gE_i(0,\vec{0})] 
 \Bigr\rangle
 \;. 
\ea
Similarly, the susceptibility $\chi^{00}$ can be written as 
\ba
 \chi^{00} & = & 
 4 \delta^{(3)}(\vec{0}) e^{-\beta M} 
 \int_{0}^{\beta}\! {\rm d}\tau \, \Bigl\langle
   \re\tr [U(\beta,\tau)U(\tau,0)] 
 \Bigr\rangle
 \nn 
 & = &
 4 \delta^{(3)}(\vec{0}) e^{-\beta M} \beta
 \Bigl\langle
   \re\tr [U(\beta,0)] 
 \Bigr\rangle
 \;. 
\ea
In total, then, 
\be
 G_E(\tau) = - \fr13 \sum_{i=1}^3 
 \frac{
  \Bigl\langle
   \re\tr \Bigl[
      U(\beta,\tau) \, gE_i(\tau,\vec{0}) \, U(\tau,0) \, gE_i(0,\vec{0})
   \Bigr] 
  \Bigr\rangle
 }{
 \Bigl\langle
   \re\tr [U(\beta,0)] 
 \Bigr\rangle
 }
 \;, \la{GE_final}
\ee
and $\kappa$ can be obtained from the corresponding spectral function
through \eq\nr{kappa_def_3}. Note that the correlation function $G_E(\tau)$ 
is {\em positive} (in a gauge with vanishing $A_0$, one can think of it as 
$
 -\partial_\tau\partial_\sigma F(\tau-\sigma)|_{\sigma = 0}
$, where $F$ is the correlation function of $A_i$).
Eq.~\nr{GE_final} is our main result. 
A related formula in Minkowski signature was given in ref.~\cite{ct}.

It is appropriate to remark that the meaning of \eq\nr{GE_final} is 
unclear in the confinement phase of pure SU($\Nc$) gauge theory, where 
the expectation value of the Polyakov loop vanishes. In this situation, 
however, there would be a flux tube which drags the heavy quark in a way 
that is quite unlike diffusion, so it need not be surprising if the result 
for a diffusion coefficient were ill-defined.

%
\section{Perturbation theory}
\la{se:pert}

The derivation of our main result, \eq\nr{GE_final}, made no use of the 
weak-coupling expansion, and is meant to be applicable everywhere in 
the deconfined phase, particularly at the phenomenologically interesting 
temperatures of a few hundred MeV. Nevertheless, to gain some understanding
on the general shape of the corresponding spectral function, we now go to 
very high temperatures, where the weak-coupling expansion is applicable. 
Our goal is to demonstrate explicitly that even in this regime, where 
spectral functions in general have more peaks and cusps than in 
a strongly-coupled regime, ours is relatively smooth.

The leading-order (free theory) behaviors of the 
correlation function in \eq\nr{GE_final} and of the 
spectral function in \eq\nr{analytic} are easily found:
\begin{eqnarray}
 G_E(\tau) & = & g^2 C_F\, \pi^2 T^4 \left[
 \frac{\cos^2(\pi \tau T)}{\sin^4(\pi \tau T)}
 +\frac{1}{3\sin^2(\pi \tau T)} \right] + \rmO(g^4) \;, \\  
 \rho(\omega) & = & \frac{g^2C_F}{6\pi} \omega^3 + \rmO(g^4)
 \;, \la{rho_free}
\end{eqnarray}
where $C_F \equiv (\Nc^2 - 1)/(2\Nc)$.
This shows that at the free level the spectral function has 
no zero-frequency peak, in contrast to the spectral functions relevant 
for transport coefficients and vector current correlators
(which have $\delta$-function peaks at this order).  
Given that $\rho(\omega)$ in \eq\nr{rho_free}
vanishes faster than $\propto \omega$, the diffusion constant
$\kappa$ of \eq\nr{kappa_def_3}
is zero; we must work harder to find the 
leading non-trivial behavior at small frequency.

At next-to-leading order, $\rmO(g^4)$, 
the intercept $\kappa$
%
becomes non-vanishing~\cite{mt}: 
\be
 \kappa = \frac{g^2 C_F T}{6\pi} \mD^2 
 \biggl(\ln\frac{2 T}{\mD} + 
 \fr12 - \gamma_\rmii{E} + \frac{\zeta'(2)}{\zeta(2)}
 + \frac{\Nf \ln 2}{2\Nc + \Nf} \biggr)
 \biggl( 1 + \rmO(g) \biggr)
 \;, \la{kappa_lo}
\ee
where $\mD^2 = g^2 T^2 (\Nc/3 + \Nf / 6)$.
As indicated, corrections to this expression start already
at $\rmO(g)$, and have in fact recently been determined~\cite{chm}.

In order to learn how ``easy'' it is to extract the intercept 
$\kappa$ in practice,  let us calculate more carefully the 
small-$\omega$ behavior of the spectral function in \eq\nr{analytic}.
We restrict, in the following, to frequencies at most of 
the order of the plasmon (or Debye) scale, $\omega\lsim gT$.
Defining $\kappa(\omega)$ to be the product on the right-hand side
of \eq\nr{kappa_def_3}, the difference $[\kappa(\omega)-\kappa]$ 
gets contributions only from soft momenta $k\sim \mD$, and can be 
calculated at tree-level using Hard Thermal Loop propagators.
Moreover, the Wilson lines in 
\eq\nr{GE_final} can be set to unity. Inserting the gluon propagator
\be
 \langle A^a_{\mu} ( x) A^b_\nu ( y) \rangle = 
 \delta^{ab} \Tint{K} e^{i K\cdot ( x -  y)}
 \biggl[
   \frac{P^T_{\mu\nu}( K)}{ K^2 + \Pi_T(K)} + 
   \frac{P^E_{\mu\nu}( K)}{ K^2 + \Pi_E(K)} + 
   \xi \frac{ K_\mu  K_\nu}{(K^2)^2} 
 \biggr] 
 \;,  \la{prop}
\ee
where $\xi$ is the gauge parameter, and carrying out the Fourier
transform in \eq\nr{tildeGE}, we get 
\be
 \tilde G_E(\omega_n) = - \frac{g^2 C_F}{3}
 \int \! \frac{{\rm d}^3\vec{k}}{(2\pi)^3}
 \biggl[
   \frac{2\omega_n^2}{\omega_n^2 + {k}^2 + \Pi_T(\omega_n,\vec{k})} +  
   \frac{\omega_n^2 + {k}^2}
    {\omega_n^2 + {k}^2 + \Pi_E(\omega_n,\vec{k})} 
 \biggr] 
 \;,
\ee
where $k\equiv |\vec{k}|$.
After the analytic continuation in \eq\nr{analytic}, 
$\omega_n \to -i [\omega + i 0^+]$, the self-energies become
(see, e.g..\ ref.~\cite{kg})
\ba
 \Pi_T(-i(\omega + i 0^+),\vec{k}) & = & 
 \frac{\mD^2}{2} 
 \biggl\{ 
   \frac{\omega^2}{{k}^2} + 
   \frac{\omega}{2 {k}}
   \biggl[
     1 -  \frac{\omega^2}{{k}^2}
   \biggr] 
   \ln\frac{\omega + i 0^+ + {k}}{\omega + i 0^+ - {k}}
 \biggr\} 
 \;, \\
 \Pi_E(-i(\omega + i 0^+),\vec{k}) & = & 
 \mD^2 
   \biggl[
     1 -  \frac{\omega^2}{{k}^2}
   \biggr] 
   \biggl[
     1 -     
     \frac{\omega}{2 {k}}
   \ln\frac{\omega + i 0^+ + {k}}{\omega + i 0^+ - {k}}
   \biggr] 
 \;.
\ea
This leads to 
Landau cut contributions at $k> \omega$, and 
plasmon pole contributions at $k < \omega$. Concretely,
\ba
 \kappa(\omega) - \kappa & = & 
 \frac{2 g^2 C_F T}{3} \times \frac{4\pi}{(2\pi)^3} \times \pi \mD^2 \times
 \biggl\{ \nn 
 & & 
 \int_{\hat\omega}^{\infty} \! {\rm d}\hat k \, \hat k^2
 \frac{2\hat \omega \times \frac{\hat\omega}{4\hat k}
 \Bigl( 1 - \frac{\hat\omega^2}{\hat k^2}\Bigr)}
 {
  \Bigl( 
    \hat k^2 - \hat\omega^2 + \fr12 
     \Bigl[ 
       \frac{\hat\omega^2}{\hat k^2} + 
       \frac{\hat\omega}{2\hat k} 
       \Bigl( 1 - \frac{\hat\omega^2}{\hat k^2}\Bigr) 
       \ln\frac{\hat k + \hat\omega}{\hat k - \hat\omega}
     \Bigr]
  \Bigr)^2
 + \Bigl( 
     \frac{\hat\omega\pi}{4\hat k}
  \Bigr)^2
       \Bigl( 1 - \frac{\hat\omega^2}{\hat k^2}\Bigr)^2 
 } \nn 
 & & + \; 
 \int_{0}^{\infty} \! {\rm d}\hat k \, \hat k^4
 \biggl[ 
 \frac{
 \theta(\hat k - \hat\omega) \times
 \frac{1}{\hat\omega}\times \frac{\hat\omega}{2\hat k}
 }
 {
  \Bigl( 
    \hat k^2 
     + 1 - 
       \frac{\hat\omega}{2\hat k} 
       \ln\frac{\hat k + \hat\omega}{\hat k - \hat\omega}
       \Bigr)^2
 + \Bigl( 
     \frac{\hat\omega\pi}{2\hat k}
  \Bigr)^2
 } 
 - \frac{\frac{1}{2\hat k}}{(\hat k^2 + 1)^2}
 \biggr]
 \nn 
 & & + \; 
 {2 \hat\omega}
 \left. 
   \frac{\hat k_T^3 (\hat\omega^2 - \hat k_T^2)}
   {|3(\hat k_T^2 - \hat\omega^2)^2 -\hat\omega^2|}
 \right|_{\hat k_T^2 - \hat\omega^2 + \fr12
      [\frac{\hat\omega^2}{\hat k_T^2}+
        \frac{\hat\omega}{2\hat k_T} 
       ( 1 - \frac{\hat\omega^2}{\hat k_T^2} ) 
       \ln\frac{\hat\omega + \hat k_T }{\hat\omega - \hat k_T}
       ] \; = \; 0 }
\nn 
 & & + \; 
 \frac{1}{\hat\omega}
 \left. 
   \frac{\hat k_E^3 (\hat\omega^2 - \hat k_E^2)}
   {|3(\hat k_E^2 - \hat\omega^2) + 1|}
 \right|_{\hat k_E^2 + 1 -
        \frac{\hat\omega}{2\hat k_E} 
       \ln\frac{\hat\omega + \hat k_E }{\hat\omega - \hat k_E} \; = \; 0 }
 \quad \biggr\} \;, \la{wdep}
\ea
where $\hat\omega \equiv \omega/\mD$ and $\hat k \equiv k/\mD$. 
The four terms correspond to the transverse cut, electric cut, 
transverse pole, and electric pole, respectively. 

\begin{figure}[t]


\centerline{%
 \epsfysize=9.0cm\epsfbox{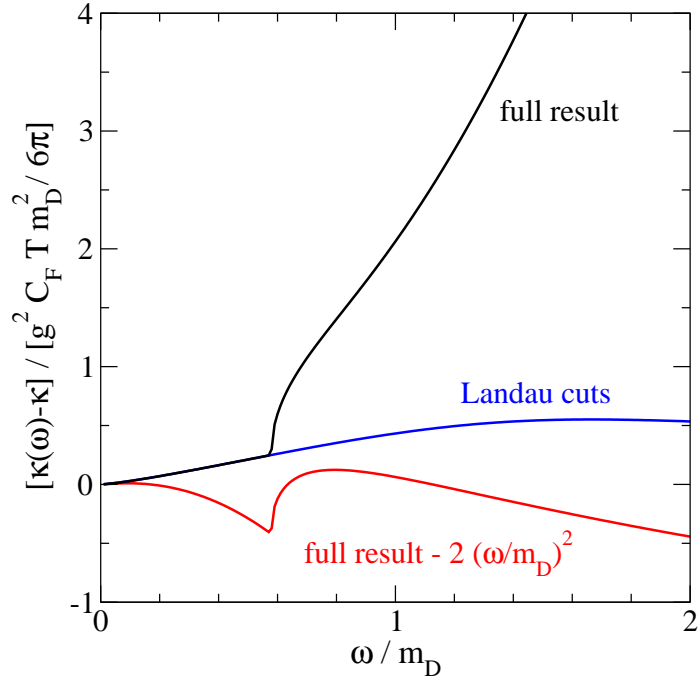}%
}

\caption[a]{\small 
  A numerical evaluation of \eq\nr{wdep},
  in units of  $g^2 C_F T \mD^2 / 6\pi$.
  The cusp is a feature of the weak-coupling
  expansion, as discussed in the text.
 }
\la{fig:wdep}
\end{figure}

The outcome of a numerical evaluation of \eq\nr{wdep} 
is plotted in \fig\ref{fig:wdep}, in units of the coefficient
$g^2 C_F T \mD^2 / 6\pi$ multiplying the logarithm in \eq\nr{kappa_lo}.
For $\hat\omega < 1/\sqrt{3}$, the result comes exclusively from the Landau
cuts; for $\hat\omega > 1/\sqrt{3}$, plasmon poles contribute as
well. For $\omega \gg \mD$, the result is dominated by 
the transverse pole, and extrapolates towards 
$\kappa(\hat\omega\gg 1)\rightarrow 
g^2 C_F T \mD^2 / 6\pi \times
2\hat\omega^2$, the free theory result.

The pattern in \fig\ref{fig:wdep} 
illustrates an important point: even at weak coupling there is 
no transport peak around the origin; 
rather $\rho(\omega)/\omega$ displays a
relatively flat behavior at $\omega\lsim \mD/\sqrt{3}$, with a significant
rise only above the Debye scale.  The only singularity
is associated with the onset of the plasmon contributions;
however this should be smoothed out in the full dynamics.
The amount of smoothening can be estimated from the (zero-momentum)
plasmon damping rate calculated in ref.~\cite{width},
$\Gamma_\rmi{pl}= 6.64 g^2\Nc T/24\pi$.  
Already for $\alpha_{\rm s}=0.05$ this gives (for $\Nc = 3, \Nf = 0 ... 4$)
a width $\Gamma_\rmi{pl}/\mD\gsim 0.2$ 
comparable to that of the cusp;
therefore we expect the true behavior
to be completely regular.
A more detailed study of the shape, 
including the effects of interactions in the small-$\omega$ regime
and ultraviolet features in the large-$\omega$ regime, is deferred
to a future publication~\cite{in_prep}. We also remark that the 
corresponding spectral functions computed for ${\mathcal N} = 4$
Super-Yang-Mills theory at infinite 't Hooft coupling show 
an analogous behavior, with the smooth infrared part ending 
in that case at $\omega\sim T$~\cite{ct,ssg}.

%
\section{Correlator in lattice regularization}
\la{se:latt}

Let us finally move to lattice regularization. In principle correlators
of the type in \eq\nr{GE_final} can be measured with standard techniques
on the lattice, in fact even at low temperatures where the signal is very 
small~\cite{kkw}. There is the problem, however, that the lattice electric 
fields require in general multiplicative renormalization factors (see, e.g., 
ref.~\cite{cm}); these depend on the details of the discretization 
chosen, and it is also not clear how they 
could be determined on the non-perturbative level\footnote{%
 For recent progress with lattice magnetic fields, 
 see ref.~\cite{gms}. 
 }. 

It appears, however, that the problem can at least be ameliorated
if we choose a discretization of the electric fields inspired by 
lattice heavy quark effective theory (see, e.g.,\ ref.~\cite{hs}). 
The spatial components of the current (\eq\nr{cJj})
could be thought of as  
\be
 \hat{\cal J}^j = 
 \frac{i}{2 a M}
 \Bigl[
  \hat\theta^\dagger({x} + \hat{j})
  U^\dagger_j(x) \hat\theta(x) 
 - 
  \hat\theta^\dagger({x})
  U_j(x) \hat\theta(x + \hat{j}) 
 - (\hat\theta\longrightarrow\hat\phi)
 \Bigr]
 \;,
\ee
where $a$ is the lattice spacing; 
$\hat{j}$ is a unit vector in the $j$-direction; 
and $U_j$ is a spatial link matrix. 
Discretizing also the time derivatives in \eq\nr{kappa_def_2}
and carrying out the contractions, we end up with a representation
of \eq\nr{GE_final} which can best be represented graphically: 
\ba
 && \hspace*{-0.5cm} \LattGE \nn
 \la{GE_lattice}
\ea
Here the direct lines within parentheses
are link matrices; reading from the right, 
the long horizontal Wilson lines in the numerator
have lengths $\tau-a$ and $\beta-\tau-a$, if the sources are placed around 
$x_0 = a/2$ and $x_0 = \tau + a/2$, respectively; 
and the denominator stands for the trace
of the Polyakov loop.
It appears that \eq\nr{GE_lattice} should be less ultraviolet 
sensitive than the usual discretizations of the electric 
fields~\cite{kkw,cm}.

Continuing with the framework of  the
lattice heavy quark effective theory, the renormalization 
of \eq\nr{GE_lattice} can also be discussed in concrete terms, 
and be related to two separate issues. First of all, 
the linearly appearing mass parameter $M$ in \eq\nr{1oM}
is no longer the pole mass but requires additive renormalization; 
second, the coefficient $c_2$ can differ from unity due to the absence
of (Euclidean) Lorentz invariance. It appears that both of these
issues could be addressed perturbatively and, in fact, even 
non-perturbatively~\cite{hs}. Since the explicit results depend on 
the particular lattice discretization chosen we do not, however, 
go into details here. 

%

%
\section{Conclusions and outlook}
\la{se:concl}

The main purpose of this paper has been to give a non-perturbative
definition to the heavy quark momentum diffusion coefficient, $\kappa$, 
allowing in principle for its lattice measurement. The basic definition
is given in terms of a certain limit of the vector current correlation
function, \eq\nr{kappa_def}. Making use of heavy quark effective theory, 
we have furthermore shown that the definition can be reduced to 
a much simpler purely gluonic correlator, given in \eq\nr{GE_final}, 
with $\kappa$ then following from \eq\nr{kappa_def_3}. 

An important consequence of these relations is that they show that 
$\kappa$ does not contain any logarithms of the heavy quark mass $M$. 
Our formulae could in principle also serve as the starting
point for a first computation of a finite-temperature 
real-time quantity
to relative accuracy $\alphas$, revealing in particular how the 
renormalization scale should be fixed.

Moving to the non-perturbative level, we have also suggested a particular
discretization of \eq\nr{GE_final}, given in \eq\nr{GE_lattice}, which 
could be free of significant renormalization factors. It remains to be 
tested in practice, however, how noisy the correlator is, 
and how fast the continuum limit can be approached. 
In addition, current practical recipes~\cite{recent} 
related to the inversion of \eq\nr{int_rel}
suffer from uncontrolled systematic uncertainties which our 
method does not remove completely, although we hope that from the practical
point of view they are less serious than in many other cases.

Assuming that 
a non-perturbative value can be obtained for $\kappa$, we can finally proceed 
to consider the thermalization rate of heavy quarks. A concrete and 
theoretically satisfactory meaning 
for a thermalization rate is provided by the heavy quark relaxation 
rate, or drag coefficient, denoted by $\eta_D$ and defined around
\eq\nr{kappa0} (the relation to thermalization follows from 
\eq\nr{p_noneq}).
Employing the fluctuation--dissipation relation
in \eq\nr{einst}, $\eta_D$ can be 
estimated as $\eta_D \simeq \kappa / 2 M T$, where $M$ is the heavy 
quark pole mass and $T$ is the temperature. Although
this relation does have ambiguities related to the definition 
of the quark mass (a pole mass has inherent non-perturbative ambiguities 
at the level of several hundreds of MeV~\cite{bb}; a treatment free of 
this problem can only be given in terms of non-perturbatively 
renormalized heavy quark effective theory~\cite{hs}), such ambiguities
should be subdominant compared with the large corrections
related to infrared sensitive thermal physics, at least
for the bottom quarks. These large thermal corrections are 
properly captured by our definition of $\kappa$, 
so $\eta_D$ should lie in the right ballpark as well. We are 
therefore very much looking forward to the first
numerical estimates of $\kappa$.

%
\section*{Acknowledgments}

We thank M.~Tassler for useful discussions. 
SCH and GDM acknowledge the hospitality of the University of Bielefeld,
where this work has been completed.
The work of SCH and GDM was supported in part by the Natural Sciences
and Engineering Research Council of Canada.  GDM thanks
the Alexander von Humboldt Foundation for its support through a 
F.\ W.\ Bessel award.



\appendix
\renewcommand{\thesection}{Appendix~\Alph{section}}
\renewcommand{\thesubsection}{\Alph{section}.\arabic{subsection}}
\renewcommand{\theequation}{\Alph{section}.\arabic{equation}}

%
\section{Examples of dynamics leading to a transport peak}

In this appendix, we review briefly two arguments through
which the Lorentzian form of the transport peak in \eq\nr{lorentz}
can be established explicitly. 

Consider first non-relativistic quantum mechanics. Let us define
$\hat v_i = \hat p_i / M_\rmi{kin}$, where $\hat p_i$ is the momentum
operator of the heavy quarks
(i.e.\ the generator of translations in their Hilbert space). 
Suppose that we have, 
through some external source field, managed to prepare a non-equilibrium state 
where there is a heavy quark with a non-zero velocity. In thermal equilibrium, 
the average velocity must vanish, so we may expect the system to behave as
\be
   \frac{{\rm d}}{{\rm d} t} \langle \hat v_i(t) \rangle_\rmi{non-eq}
  =  
 -\eta_D \, \langle \hat v_i(t) \rangle_\rmi{non-eq}
 + \rmO\Bigl(\langle \hat v_i(t) \rangle_\rmi{non-eq}^2\Bigr)
 \;.  \la{p_noneq}
\ee  
Once $t$ is so large that  
$
 \langle \hat v_i(t) \rangle_\rmi{non-eq}  \sim
 [ \langle \hat v_i^2 \rangle_\rmi{eq} ]^{1/2}
$, 
Brownian motion sets in, and the system effectively 
equilibrates. In equilibrium we may define the correlator
\be
 \Delta_{ii}(t) \equiv \left\langle \fr12 \{ \hat v_i(t), \hat v_i(0) \}
 \right\rangle_\rmi{eq}  \;. \la{p_eq}
\ee
This is an even function of $t$ and must vanish for $t\to \infty$; 
in fact, at least on certain time scales, 
it can be argued that it vanishes with the {\em same}
exponent as the non-equilibrium correlator in \eq\nr{p_noneq}
(see, e.g., \S118 of ref.~\cite{ll}): 
\be
 \Delta_{ii}(t) \stackrel{|t| \gg \beta}{\simeq} 
 \bar \Delta_{ii}\, e^{-\eta_D|t|}
 \;, \la{Delta_t}
\ee
where $\bar \Delta_{ii}$ is a constant. 
Taking a Fourier transform
yields
\be
 \tilde \Delta_{ii}(\omega) \equiv 
 \int_{-\infty}^{\infty} \! {\rm d}t \, e^{i\omega t} \, 
 \Delta_{ii}(t) 
 \stackrel{|\omega| \ll T}{\simeq} 
 \bar \Delta_{ii} \frac{2\eta_D}{\omega^2 + \eta_D^2}
 \;, 
\ee
and making use of the general relation
$
 \tilde \Delta_{ii}(\omega) =  
 \left[
 1 + 2 \nB{}(\omega) 
 \right] \rho_{ii}(\omega)
$  (see, e.g., ref.~\cite{kg}), 
where $\nB{}(\omega)\equiv 1/[\exp(\beta\omega)-1]$ and 
$\rho_{ii}(\omega)$ is the spectral function,  
we arrive at
\be
 \frac{\rho_{ii}(\omega)}{\omega}  \stackrel{|\omega|\ll T}{\approx} 
 \frac{1}{2 T} \tilde\Delta_{ii}(\omega) 
 \stackrel{|\omega|\ll T}{\simeq} 
 \bar \Delta_{ii} \frac{\beta \eta_D}{\omega^2 + \eta_D^2}
 \;. \la{pre_shape}
\ee
This indeed agrees with the functional form of \eq\nr{lorentz}.

Another example is given by classical Langevin dynamics 
(see also ref.~\cite{pt}). Essentially, we 
replace 
$
 \langle \hat p_i(t) \rangle_\rmi{non-eq} \to 
 p_i(t)
$, 
and assume the dynamics to be contained in
\ba
   \dot{p_i}(t) 
 &  = &   
 -\eta_D \, p_i(t) + \xi_i(t) 
 \;, \la{Lan1} \\ 
 \langle\!\langle \xi_i(t) \xi_j(t') \rangle\!\rangle & = & 
 \kappa_\rmi{cl} \, \delta_{ij} \delta(t-t')
 \;, \qquad
 \langle\!\langle \xi_i(t) \rangle\!\rangle = 0 
 \;, \la{Lan2}
\ea
with $\xi$ a Gaussian stochastic noise field, and 
$\langle\!\langle ... \rangle\!\rangle$ denoting an average over the noise. 
For a heavy particle the Gaussian nature
follows from the central limit theorem and the slow time scale
of its dynamics, while
the auto-correlator
$
 \kappa_\rmi{cl} = \int_{-\infty}^{\infty} \! {\rm d}t \, 
 \langle\!\langle \xi_i(t) \xi_i(0) \rangle\!\rangle
$
can be chosen such as to match that of the underlying theory, \eq\nr{kappa0}.
It is easy to verify that within this dynamics, for a distribution
with density $T\chi^{00}$ of heavy quarks, the equilibrium
correlator is exactly \eq\nr{lorentz}.

\newpage


\end{document}